\documentclass[12pt]{elsarticle}
\usepackage{graphicx}
\usepackage{refmerge}
\usepackage{epsfig}
\usepackage{lineno}
\begin{document}
\date{\today}

\title{\bf{ \boldmath
STUDY OF THE PROCESS $e^+e^-\to p\bar{p}$ IN THE C.M. ENERGY RANGE 
FROM THRESHOLD TO 2 GEV WITH THE CMD-3 DETECTOR
}}

\author[adr1,adr2]{R.R.Akhmetshin}
\author[adr1,adr2]{A.N.Amirkhanov}
\author[adr1,adr2]{A.V.Anisenkov}
\author[adr1,adr2]{V.M.Aulchenko}
\author[adr1]{V.Sh.Banzarov}
\author[adr1]{N.S.Bashtovoy}
\author[adr1,adr2]{D.E.Berkaev}
\author[adr1,adr2]{A.E.Bondar}
\author[adr1]{A.V.Bragin}
\author[adr1,adr2]{S.I.Eidelman}
\author[adr1,adr6]{D.A.Epifanov}
\author[adr1,adr3]{L.B.Epshteyn}
\author[adr1,adr2]{A.L.Erofeev}
\author[adr1,adr2]{G.V.Fedotovich}
\author[adr1,adr2]{S.E.Gayazov}
\author[adr1,adr2]{A.A.Grebenuk}
\author[adr1,adr2]{S.S.Gribanov}
\author[adr1,adr2,adr3]{D.N.Grigoriev}
\author[adr1,adr2]{E.M.Gromov}
\author[adr1]{F.V.Ignatov}
\author[adr1,adr2]{V.L.Ivanov}
\author[adr1]{S.V.Karpov}
\author[adr1]{A.S.Kasaev}
\author[adr1,adr2]{V.F.Kazanin}
\author[adr1,adr2]{\fbox{B.I.Khazin}}
\author[adr1]{A.N.Kirpotin}
\author[adr1,adr2]{I.A.Koop}
\author[adr1,adr2]{O.A.Kovalenko}
\author[adr1]{A.N.Kozyrev}
\author[adr1,adr2]{E.A.Kozyrev}
\author[adr1,adr2]{P.P.Krokovny}
\author[adr1,adr3]{A.E.Kuzmenko}
\author[adr1,adr2]{A.S.Kuzmin}
\author[adr1,adr2]{I.B.Logashenko}
\author[adr1,adr2]{P.A.Lukin}
\author[adr1]{K.Yu.Mikhailov}
\author[adr1]{V.S.Okhapkin}
\author[adr1]{A.V.Otboev}
\author[adr1]{Yu.N.Pestov}
\author[adr1]{S.G.Pivovarov}
\author[adr1,adr2]{A.S.Popov\fnref{tnot}}
\author[adr1,adr2]{G.P.Razuvaev}
\author[adr1]{A.L.Romanov}
\author[adr1,adr2]{A.A.Ruban}
\author[adr1]{N.M.Ryskulov}
\author[adr1,adr2]{A.E.Ryzhenenkov}
\author[adr1,adr2]{V.E.Shebalin}
\author[adr1,adr2]{D.N.Shemyakin}
\author[adr1,adr2]{B.A.Shwartz}
\author[adr1,adr2]{D.B.Shwartz}
\author[adr1,adr4]{A.L.Sibidanov}
\author[adr1]{P.Yu.Shatunov}
\author[adr1]{Yu.M.Shatunov}
\author[adr1,adr2]{E.P.Solodov}
\author[adr1]{V.M.Titov}
\author[adr1,adr2]{A.A.Talyshev}
\author[adr1]{A.I.Vorobiov}
\author[adr1]{Yu.V.Yudin}
\author[adr1]{Yu.M.Zharinov}

\address[adr1]{Budker Institute of Nuclear Physics, SB RAS, 
Novosibirsk, 630090, Russia}
\address[adr2]{Novosibirsk State University, Novosibirsk, 630090, Russia}
\address[adr3]{Novosibirsk State Technical University, 
Novosibirsk, 630092, Russia}
\address[adr4]{Department of Physics and Astronomy, P.O. Box 3055 Victoria, B.C.,
CANADA, V8W 3P6}
\address[adr6]{University of Tokyo, Department of Physics, 
7-3-1 Hongo Bunkyo-ku Tokyo, 113-0033, Japan}

\fntext[tnot]{\vspace {0.5cm}Corresponding author:aspopov1@inp.nsk.su}


\vspace{0.7cm}

\begin{abstract}
\hspace*{\parindent}
Using a data sample of 6.8 pb$^{-1}$  collected 
with the CMD-3 detector at the VEPP-2000 $e^+e^-$ collider 
we select about 2700 events of  the process $e^+e^- \to  p\bar{p}$  
and measure its cross section at 12 energy points with about 6\% 
systematic uncertainty. 
From the  angular distribution of produced nucleons we obtain the ratio 
$G_E/G_M$.

\end{abstract}

\maketitle

\baselineskip=17pt

\section{ \boldmath INTRODUCTION}
\hspace*{\parindent}

\begin{figure}[ptb]
\begin{center}
\includegraphics[width=0.4\textwidth] {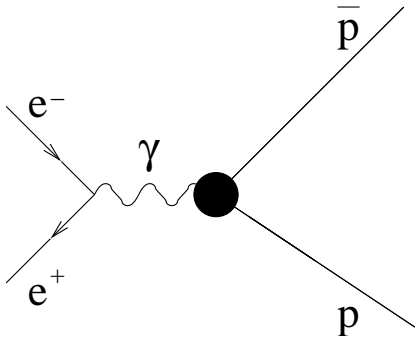}
\caption{The Feynman diagram of the process $e^+e^- \to p\bar{p}$~\label{fig:SimplePP}.}
\end{center}
\end{figure}

The Born cross section of the process $e^+e^- \to p\bar{p}$ shown 
in Fig.~\ref{fig:SimplePP} is given by 

\begin{equation}
\sigma_{p\bar{p}}(s) = \frac{4\pi\alpha^{2}\beta C}{3s} \left [|G_M(s)|^{2} + \frac{2M_p^2}{s}|G_E(s)|^{2}\right],
\label{eq3}
\end{equation}
\noindent
where $\sqrt{s} = 2E_{\rm beam} = E_{\rm c.m.}$ is the center-of-mass energy, 
$M_p$ is the proton mass, and $\beta = \sqrt{1-4M_p^2/s}$. The 
Sommerfeld-Gamov-Sakharov factor~\cite{Coulomb}
 $C= y/(1-e^{-y})$, $ y = \pi\alpha/\beta$, takes into account the 
Coulomb final state interaction. The cross section depends on the electric 
($G_{E}$) and magnetic ($G_{M}$) form factors, which are equal at the  
threshold. To compare different experiments, the effective form factor
\begin{equation}
|F(s)|^{2} = \frac{|G_{M}|^{2}+\frac{2M_{p}^{2}}{s}|G_{E}|^{2}}{1+\frac{2M_{p}^{2}}{s}}
\label{form}
\end{equation}
is usually defined. 

In early experiments at the electron-positron colliders~\cite{DM1, DM2, fenice, bes} 
in the energy range between the proton-antiproton threshold 
and $E_{c.m.}$= 2 GeV,
the cross section of the process $e^+e^- \to p\bar{p}$ has been measured
at six energy points only. The accuracy of these measurements is about 25-30\% 
and the $|G_{E}|=|G_{M}|$ assumption is made. In the PS170 experiment at 
LEAR~\cite{lear} the measurement of the effective proton form factor 
and the first measurement of the $|G_{E}/G_{M}|$ ratio have been performed in 
the process $p\bar{p} \to e^+e^-$  with 30\% accuracy. The most accurate 
measurements of the $e^+e^- \to p\bar{p}$ cross section, 
the effective form factor, and the $|G_{E}/G_{M}|$ ratio have been performed 
with the BaBar~\cite{babar} detector using the initial-state radiation (ISR) 
method. However, PS170 and BaBar results contradict to each other, and 
new experiments are obviously required.

Additional interest to this energy range is related to an unusual behavior 
of the $e^+ e^- \to 3(\pi^+\pi^-)$ cross section~\cite{Solodov} near 
the proton-antiproton threshold.

\section{ \boldmath THE CMD-3 DETECTOR}
\hspace*{\parindent}

The CMD-3 detector~\cite{cite_cmd3_1,cite_cmd3_2} is installed in one of the two interaction regions at the electron-positron collider VEPP-2000~\cite{cite_vepp_2}. The design luminosity of the VEPP-2000 is $10^{32}$ cm$^{-2}$s$^{-1}$ at the maximum center-of-mass energy $E_{\rm c.m.}$ = 2 GeV. The detector tracking system consists of the cylindrical drift chamber (DC) and double-layer cylindrical multiwire proportional Z-chamber, both used for a trigger, and both installed 
inside a thin (0.2 $X_{0}$) superconducting solenoid with 1.3 T field. 
The beam pipe inside the DC is made of 0.5 mm aluminum with 17 mm inner radius. An inner shell of DC is made of a carbon-fiber-reinforced polymer (CFRP) and has 20 mm radius.
The DC contains 1218 hexagonal cells and allows to measure charged particle momentum with 1.5-4.5$\%$ accuracy in the 100-1000 MeV/c range, and provides the measurement 
of the polar ($\theta$) and azimuthal ($\phi$) angles with 20 mrad and 
3.5-8.0 mrad accuracy, respectively. An amplitude information from the DC 
wires is used to measure ionization losses of charged particles with $ {\sigma}_{dE/dx}=$11-14\% accuracy. Two  electromagnetic calorimeters (a  
liquid xenon (LXe) one with 5.4 $X_{0}$ and CsI crystals with 8.1 $X_{0}$) 
are placed in the barrel outside the solenoid.  BGO crystals with 13.4 $X_{0}$ are used as the end-cap calorimeters. The return yoke of the detector is surrounded by scintillation counters, which are used to veto cosmic events.

We use the data samples of 2011 (1.0 T field) and 2012 (1.3 T field) runs, collected at twelve beam energy points for an integrated luminosity of 6.8 pb$^{-1}$.
To study the detector response to a proton-antiproton pair and 
determine the detection efficiency, we have developed a Monte Carlo (MC) 
simulation of our detector based on the GEANT4~\cite{GEANT4} package, 
and all simulated events pass the reconstruction and selection procedures. The MC simulation includes soft photon radiation by initial electron or positron, calculated according to Ref.~\cite{Sibid}.

\begin{figure}[ptb]
\begin{minipage}[t]{0.45\textwidth}
\centerline{\includegraphics[width=0.95\textwidth,keepaspectratio]{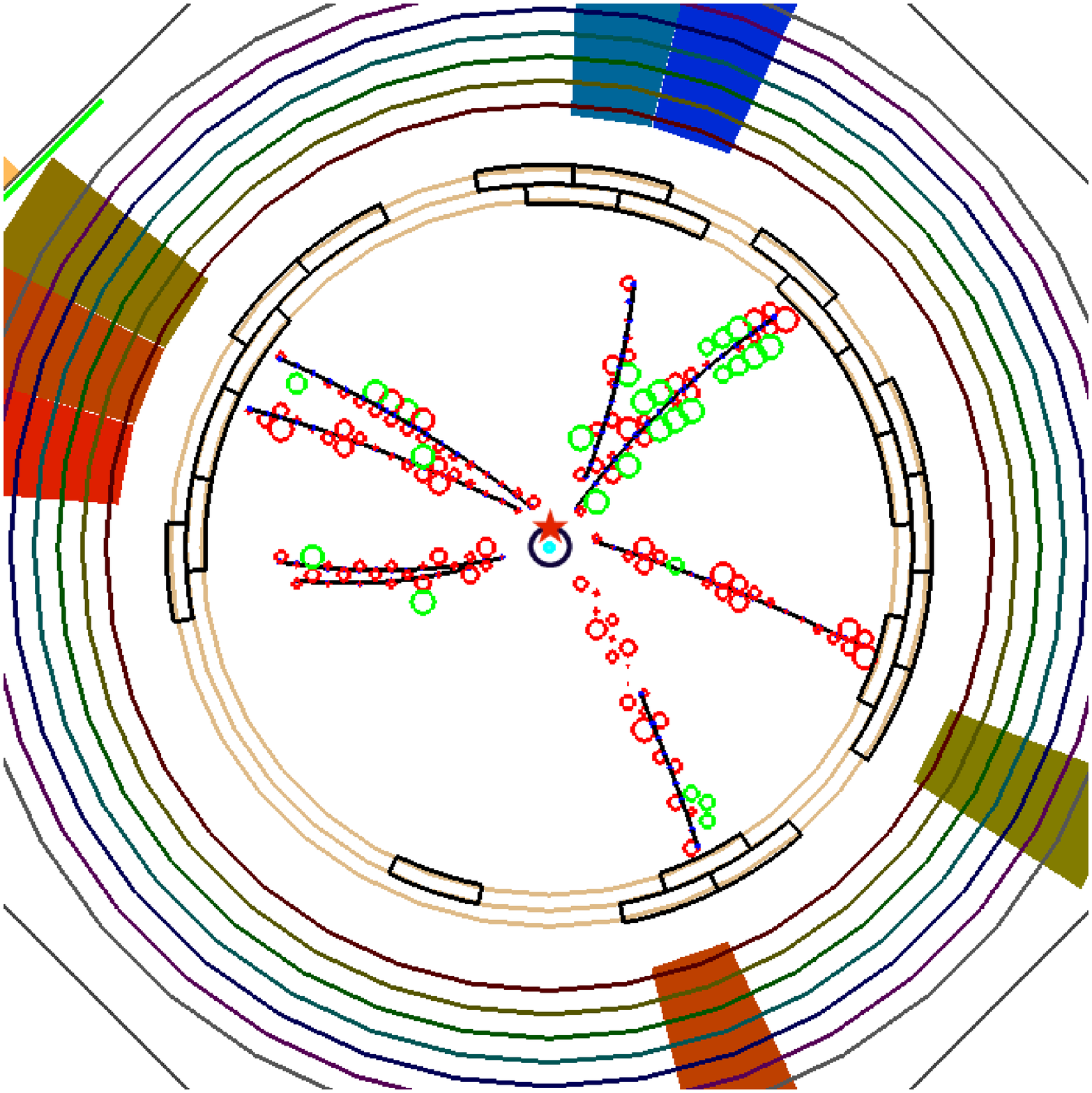}}
\caption{Example of an $e^+ e^- \to p\bar{p}$ event at $E_{\rm beam}=945$ MeV. 
The antiproton stops and annihilates in the beam pipe with production of several secondary particles.\label{fig:ev_below}}
\end{minipage}
\hfill
\begin{minipage}[t]{0.45\textwidth}
\centerline{\includegraphics[width=0.95\textwidth,keepaspectratio]{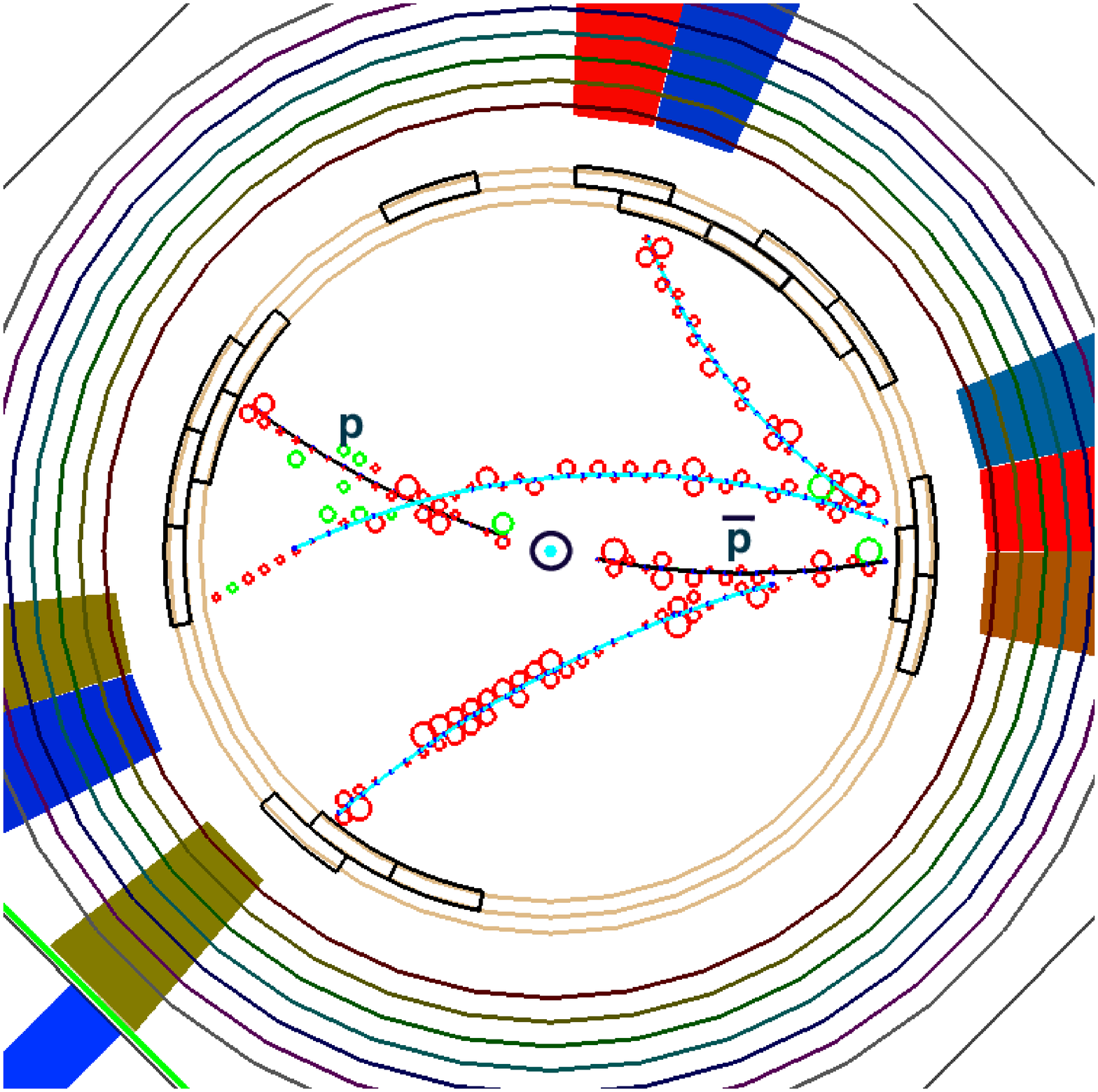}}
\caption{Example of an $e^+ e^- \to p\bar{p}$ event at $E_{\rm beam}=970$ MeV. The proton is absorbed in the Z-chamber and the antiproton annihilates with production of several secondary particles three of which come back to the DC.\label{fig:event_above}}
\end{minipage}\hspace{0.5cm}
\end{figure}

\section{ \boldmath EVENT SELECTION}

\hspace*{\parindent}
In the studied energy range nucleons have low velocity and high 
ionization losses. 
According to MC simulation, when the beam energy is less than 950 MeV, all protons and antiprotons stop in the beam pipe or in the inner DC shell, and antiprotons annihilate producing several secondary particles. 
An example of such an event is shown in Fig.~\ref{fig:ev_below}.
When $E_{\rm c.m.}$ is above 952 MeV, almost all nucleons reach 
the DC sensitive volume, and stop in the DC outer shell or in the Z-chamber. 
Figure~\ref{fig:event_above} shows an example of an event, when both proton and antiproton are detected in the DC.
For beam energies between 950 and 952 MeV, only part of nucleons 
penetrate into the DC volume. 
According to these differences in the nucleon path to annihilation, 
we use two different  approaches for the signal selection.

\subsection{Nucleons reach the DC sensitive volume}
\label{sec:coll}

The selection criteria for this class of events are the following:

a) There are two opposite-charge tracks with the number of DC hits 
$N_{\rm hit} >$ 4. They are collinear 
(($\delta\theta <$ 0.25 rad \& $\delta\phi <$ 0.15 rad) or ($\delta\theta <$ 0.4 rad \& $\delta\phi <$ 0.5 rad) for $E_{\rm beam}<$955 MeV);
 For $E_{\rm beam}<$955 MeV  we have very soft  momentum of $P\bar P$ particles with high multiple scattering.   

b) The tracks are originating from the beam interaction region within 10 cm 
along the beam axis and within 1 cm in the transverse direction.

c) Momenta of both tracks are close to each other $|p_{1}-p_{2}|/|p_{1}+p_{2}| <$ 0.15 ($<$ 0.5 for $E_{\rm beam}<$955 MeV);

d) The total energy deposition in the calorimeters is more than 200 MeV.

\begin{figure}[ptb]
\centerline{\includegraphics[width=0.5\textwidth,keepaspectratio]{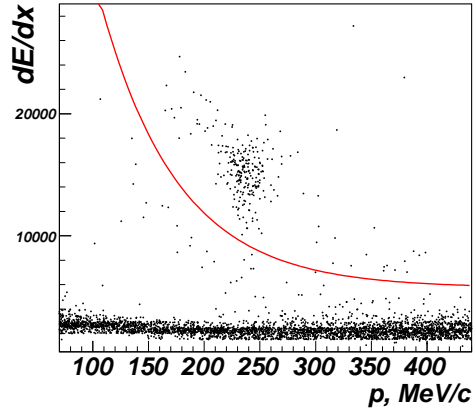}}
\caption{ The dE/dx vs momentum distribution for tracks at $E_{\rm beam}$=970 MeV. The line shows the applied selection.} \label{fig:dedx}
\end{figure}

Figure~\ref{fig:dedx} shows a scatter plot of the ionization losses ($dE/dx$) in DC vs momentum for a selected pair of tracks. A signal from $p\bar{p}$ events is clearly seen. We require both tracks to have ionization losses above a value, which is calculated by taking into account the average $dE/dx$ value and $dE/dx$ resolution at the  measured momentum. The line in Fig.~\ref{fig:dedx} shows the applied selection.

\begin{figure}[ptb]
\begin{minipage}[t]{0.45\textwidth}
\centerline{\includegraphics[width=1.2\textwidth,keepaspectratio]{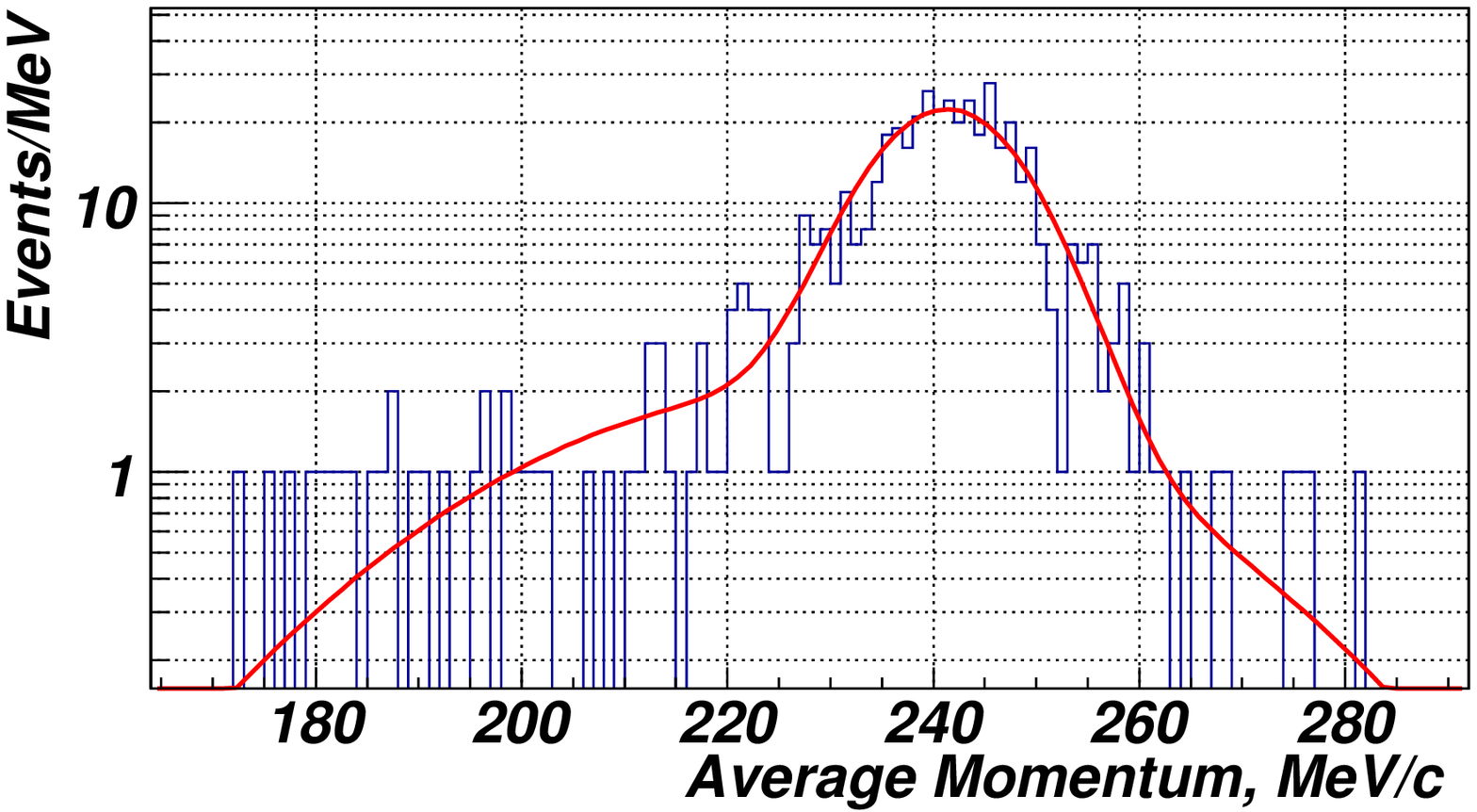}}
\caption{Average of absolute momentum values of proton and antiproton for data at $E_{\rm beam}$=970 MeV (histogram). The line shows a fit described in the text.\label{fig:moment}}
\end{minipage}
\hfill
\begin{minipage}[t]{0.45\textwidth}
\centerline{\includegraphics[width=1.2\textwidth]{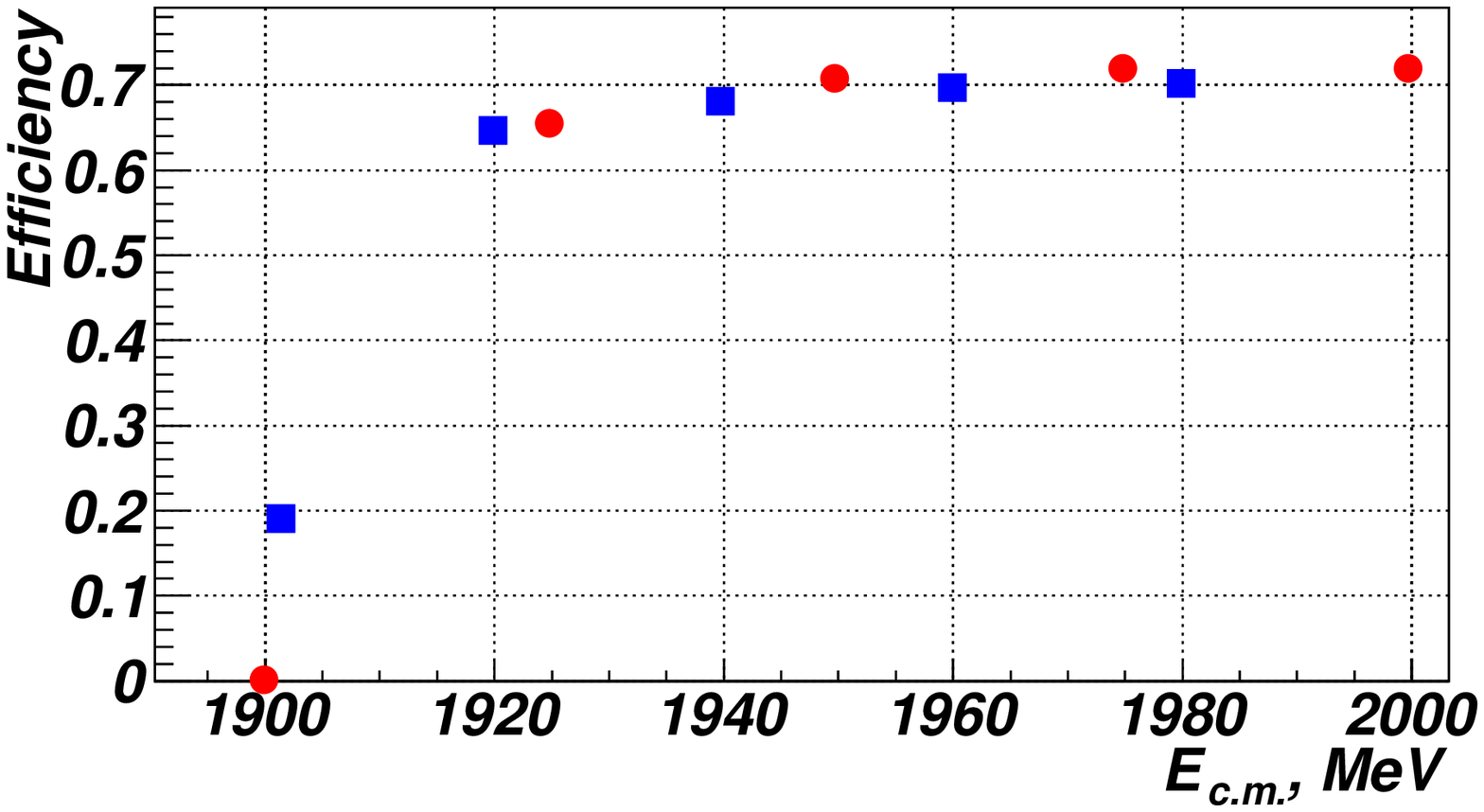}}
\caption{The detection efficiency of collinear $p\bar{p}$ pairs vs $E_{\rm c.m.}$ for the 2011 (circles) and 2012 run (squares).\label{fig:eff_above}}
\end{minipage}
\end{figure}

The distribution of the average absolute value of the nucleon momentum  
for selected $p\bar{p}$ events at $E_{\rm beam}$=970 MeV is shown in Fig.~\ref{fig:moment}. The number of background events in the signal range is expected to be negligible.  The left tail 
of the distribution is mainly due to the initial state radiation resulting in nucleons with a smaller momentum and higher $dE/dx$ value.

 The ``declared'' collider beam energy was not very precise, and for a few energy points the energy was continuously monitored during data taking, using the Back-Scattering-Laser-Light system~\cite{laser}. Based on these measurements and comparing the average momentum value for data and simulation, we can determine the c.m. energy with better accuracy.

We simulate the process  $e^+ e^- \to p\bar{p}$ at the ``declared'' collider beam energies, apply the above selections, and fit the 
average momentum distribution of the proton-antiproton pairs with a sum of two Gaussian functions. The experimental distribution at each energy point is fitted with the corresponding MC-simulated function, convolved with an additional normal distribution, which takes into account a data-MC difference in the  detector resolution. The number of events, variance of the additional normal distribution, and the momentum difference between simulation and experiment are floating. The fit curve is demonstrated in Fig.~\ref{fig:moment} by the line. Using the obtained momentum difference and proton mass value we calculate a beam energy 
shift $E_{\rm beam}^{\rm shift}$, listed in Table~\ref{table:sec_above} for each energy point.

 The detection efficiency $\epsilon_{\rm coll}$ for this class of events is calculated as  the ratio of the number of selected $p\bar{p}$ pairs to that of 
all MC-simulated events.
The energy dependence of $\epsilon_{coll}$ is shown in Fig.~\ref{fig:eff_above} for two experimental runs. 

To estimate a data-MC difference in the detection efficiency, 
we select a  pure class of events  with a detected antiproton and check how often we reconstruct the opposite proton. We use the following selection criteria:

- one or two tracks coming from the beam interaction region within 10 cm along the beam axis and within 1 cm in the transverse direction.

- one of these tracks has negative charge, has the number of hits $N_{\rm hit} > $9 with high ionization losses in DC ($\frac{dE}{dx}>\frac{dE}{dx}_{p}^{\rm mean}-\sigma_{\frac{dE}{dx}}$), and associated energy deposition in the calorimeters is more than 100 MeV;

- the total energy deposition in the calorimeters is from 300 to 1100 MeV.

\begin{figure}[ptb]
\begin{minipage}[t]{0.49\textwidth}
\centerline{\includegraphics[width=0.99\textwidth,keepaspectratio]{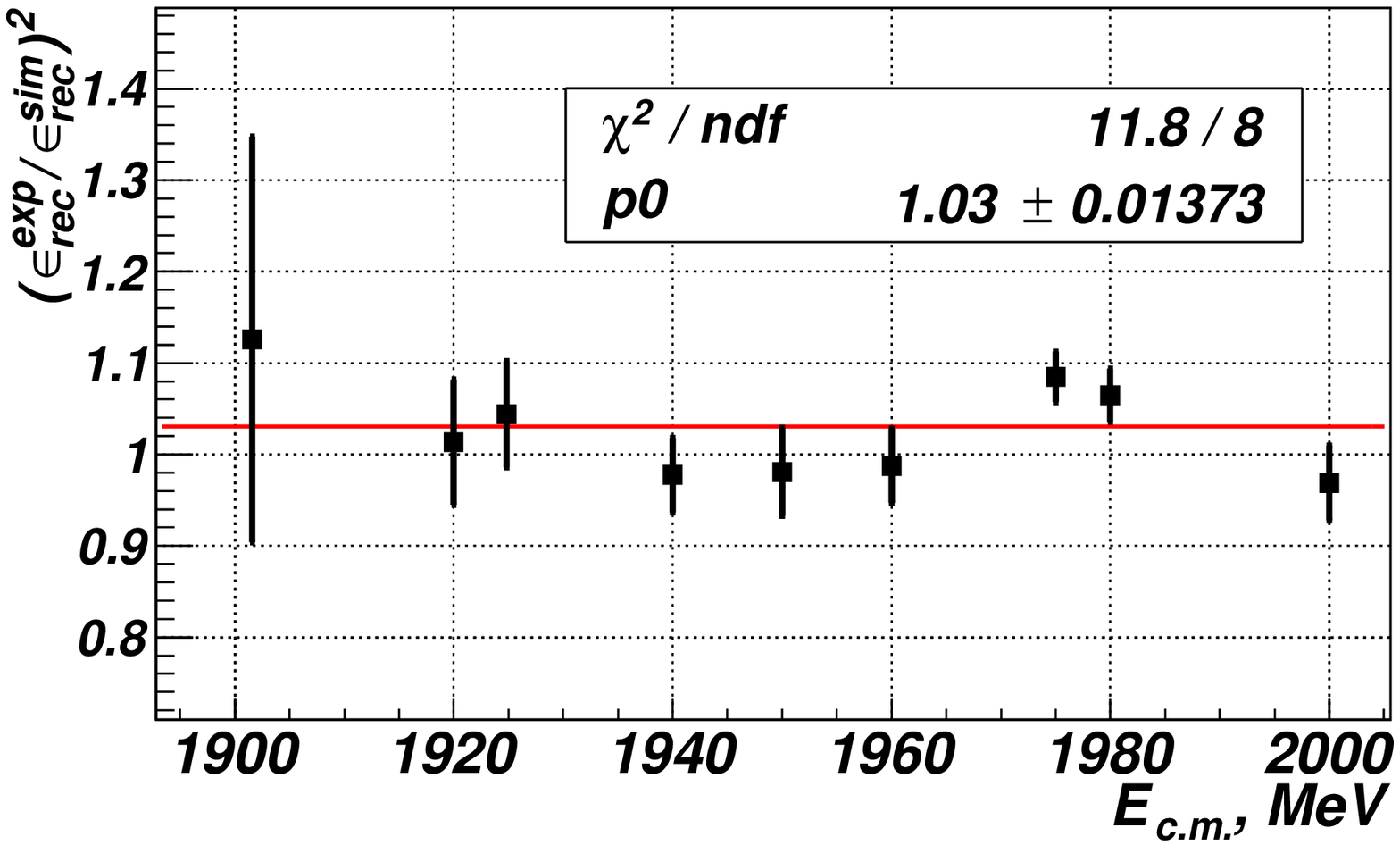}}
\caption{The correction due to the data-MC difference in the proton/antiproton 
detection efficiency. \label{fig:eff_reg}}
\end{minipage}
\begin{minipage}[t]{0.49\textwidth}
\centerline{\includegraphics[width=0.99\textwidth,keepaspectratio]{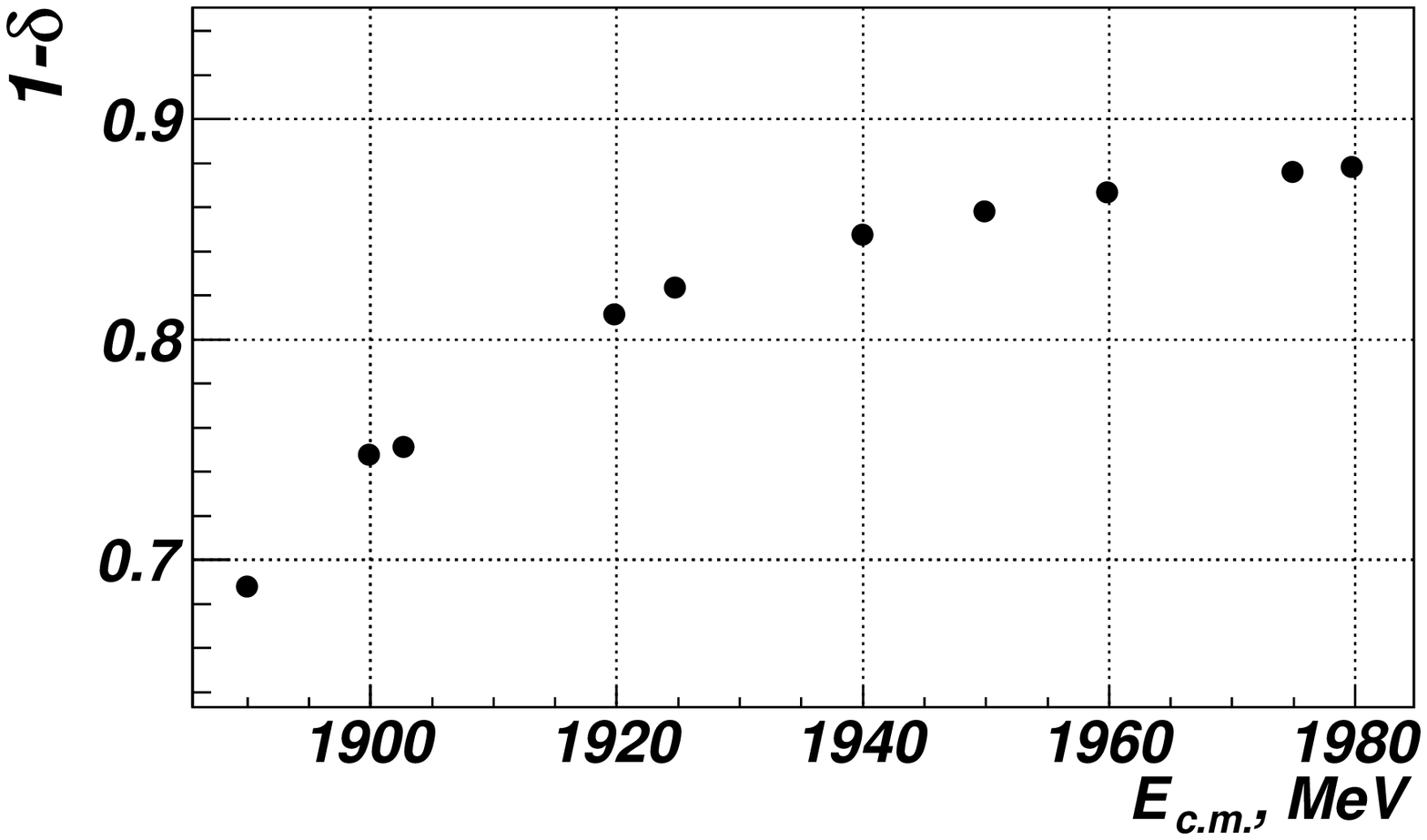}}
\caption{The radiative corrections vs c.m. energy. \label{fig:radcorr}}
\end{minipage}
\end{figure}

Using these selections we obtain the proton detection efficiency for data $\epsilon_{\rm reg}^{\rm exp}$ and MC-simulation  $\epsilon_{\rm reg}^{\rm sim}$, which are calculated as a ratio of the number of events with found protons to that of all events with antiprotons. Because of the large proton background from beam-gas interactions we cannot select a pure sample of detected protons, and assume equal detection efficiencies for protons and antiprotons. 
The squared ratio of efficiencies found for data and MC-simulation 
is shown in Fig.~\ref{fig:eff_reg} vs c.m. energy, and gives an estimate of data-MC difference. 
We fit these points with a constant and obtain the value  
$R = 1.030 \pm 0.014$ close to unity and it  is used as an estimate of the systematic error on the detection efficiency.
At the point $E_{\rm c.m.} = 1901.6$~MeV we use the correction as determined above, because the effect of material and thickness uncertainties 
of the beam pipe is too big.  
 
 At each energy point the $e^+e^- \to p\bar{p}$ cross section for this class of events is calculated from 
\begin{equation}
\sigma_{\rm Born}=\frac{N_{p\bar{p}}}{L\,\epsilon_{\rm coll}(1-\delta)R},
\label{sigma_above}
\end{equation}
where $L$ is the integrated luminosity, and $\epsilon_{\rm coll}$ is the detection efficiency.
Figure~\ref{fig:radcorr} shows energy dependence of the radiative correction (1-$\delta$) calculated according to Ref.~\cite{Sibid}.

The c.m. energy , beam energy shift, luminosity, number of selected $e^+e^- \to p\bar{p}$ events, detection efficiency, radiative correction, and cross section
are listed in Table~\ref{table:sec_above}.

\subsection{Nucleons are absorbed in the beam pipe or the DC inner shell}

When an antiproton stops in the material of the beam pipe or in the DC 
inner shell, it annihilates with production of several secondary particles, which are mostly pions. Part of the produced negative pions are captured by nuclei and induce nucleus fragmentation and production of protons and neutrons as well as deuterons and tritons. 

Candidates to this class of events are selected with the following criteria:

a) an event has a vertex with 4 or more tracks located in front of the beam pipe or the DC inner shell;

b) an event has no tracks with energy deposition in calorimeters higher than 400 MeV.

We verify these criteria in a special run without beams, and conclude that cosmic events are completely rejected by these criteria. 

The main remaining background is due to the interactions of the 
particles lost from the beams  with the detector material. Several pions, 
protons and heavier particles are produced in such interactions. We study 
this background in a  special run with one electron or positron beam only, 
and using data from c.m. energy points below the threshold.  

\begin{figure}[ptb]
\begin{minipage}[t]{0.99\textwidth}
\centerline{\includegraphics[width=0.99\textwidth,keepaspectratio]{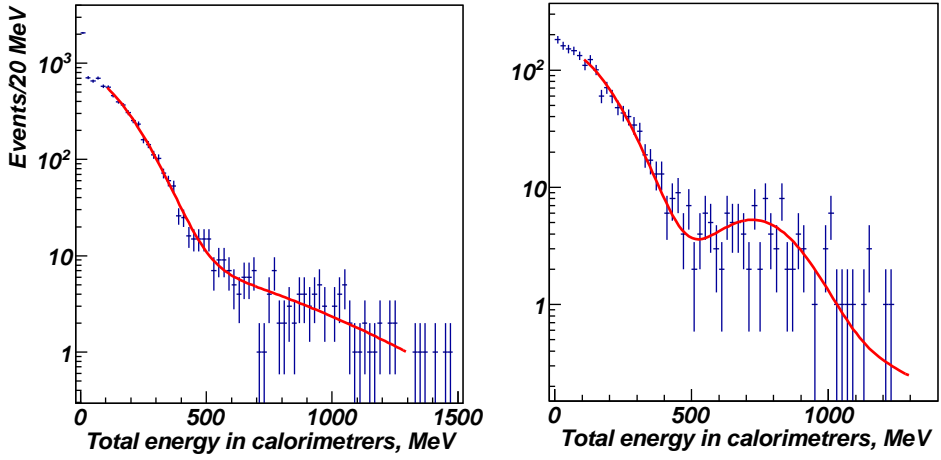}}
\caption{The distribution of total energy deposition in calorimeters for candidates to $e^+ e^- \to p\bar{p}$ events with c.m. energy below threshold (left) and above threshold (right)\label{fig:en}}
\end{minipage}
\end{figure}

\begin{figure}[htb]
\begin{center}
\centerline{\includegraphics[width=0.6\textwidth,keepaspectratio]{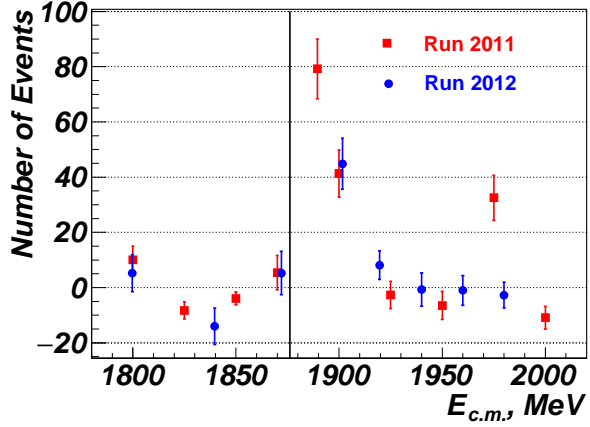}}
\caption{Numbers of $e^+ e^- \to p\bar{p}$ events with annihilation in the beam pipe and in the DC inner shell. The vertical line shows the proton-antiproton production threshold. \label{fig:nevbelow}}
\end{center}
\end{figure}

To obtain the number of $e^+e^- \to p\bar{p}$ events we use additional 
information from the calorimeters. The distribution of the energy deposition from the background events, shown in Fig.~\ref{fig:en}(left) with a fit function, is obtained from the runs where the c.m. energy was below the $p\bar{p}$ production threshold. 
Figure~\ref{fig:en}(right) shows the combined distribution of the total 
energy deposition for three c.m. energies above threshold; 945, 950 MeV(run 2011) and 950 MeV(run 2012).
A signal from the antiproton annihilation is clearly seen. We describe this 
signal with an additional Gaussian function and use its parameters 
to obtain the number of $e^+e^- \to p\bar{p}$ events at each c.m. energy point. 
The obtained numbers of signal events vs c.m. energy are shown in Fig.~\ref{fig:nevbelow}.

To obtain the cross section we need to determine the detection efficiency for this class of events.

\begin{figure}[ptb]
 \begin{center}
 \centerline{\includegraphics[width=0.6\textwidth,keepaspectratio]{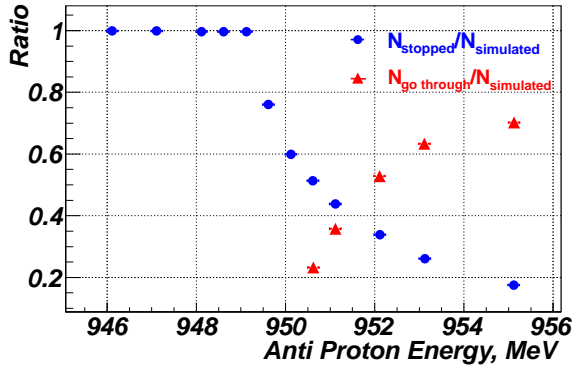}}
\caption{Data from simulation for the fraction of antiprotons annihilating at rest in the beam pipe and in the DC inner shell (dots), and those, penetrating deeper than 15 cm in DC (stars). \label{fig:stopped}}
 \end{center}
 \end{figure}

To calculate the efficiency for stopped nucleons we use our data
at the energy point 950 MeV (run 2012), where part of antiprotons stop in the beam pipe and in the DC inner shell, and annihilate, $N_{\rm ann}$, while part of $p\bar{p}$ pairs pass the DC sensitive volume, and are identified as collinear events, $N_{\rm coll}$, allowing to calculate the cross section. A  
fraction of stopped and annihilated antiprotons vs beam energy $\epsilon_{stopped}$ was obtained from simulation and   
 is shown in Fig.~\ref{fig:stopped} by circles.
Using the measured beam energy ( $E_{\rm c.m.}=950.8$ MeV see above ), we obtain 
$\epsilon_{\rm stopped} = 0.5 \pm 0.1$ with the corresponding $\epsilon_{\rm coll} = 0.20\pm0.04$ for the detected $p\bar{p}$ collinear events (Fig.~\ref{fig:eff_above}).
(The uncertainties of $\epsilon_{\rm stopped}$ was obtained from simulation with different thickness of vacuum pipe, the thickness is $0.50\pm0.05$ mm.)
At this energy point we calculate a ``visible'' cross section for the collinear events, $\sigma_{\rm vis} = N_{\rm coll}/(\epsilon_{\rm coll} L)$,  and assuming  
the same production cross section, we determine a detection efficiency 
for the annihilated antiprotons, $\epsilon_{\rm ann}$:

 \[ \epsilon_{\rm ann}\cdot \epsilon_{\rm stopped}=\frac{N_{\rm ann}}{L\cdot \sigma_{\rm vis}}=\frac{N_{\rm ann}\,\epsilon_{\rm coll} L  }{L\cdot N_{\rm coll}}=\frac{N_{\rm ann}\,\epsilon_{\rm coll}}{N_{\rm coll}}\,\,\to\,\]
\[ \to\,\epsilon_{\rm ann}= \frac{N_{\rm ann}\,\epsilon_{\rm coll}}{N_{\rm coll} \epsilon_{\rm stopped}}=\frac{(44.8\pm9.2)\cdot (0.20\pm0.04)}{(164.6\pm13.0)\cdot (0.5\pm0.1)} = 0.112\pm0.033 \]

The detection efficiency for annihilated antiprotons is found to be 
$\epsilon_{\rm ann}=0.112\pm0.033$, and because antiprotons annihilate at rest, it does not depend on the beam energy. In the 2011 run the real beam energy 
at 950 MeV is a little below the threshold of DC penetration, 
and we can not use the procedure.  At other energy points the cross section can be calculated as:

\[ \sigma_{\rm Born}=\frac{N_{p\bar{p}}}{L\epsilon_{\rm ann}\epsilon_{\rm stopped}(1-\delta)}.\]

The obtained values are listed in Table~\ref{table:sec_below}.

\section{The $|G_E/G_M|$ ratio}

The $e^+e^- \to p\bar{p}$ cross section depends on the proton/antiproton polar angle $\theta$ as:

\begin{equation}
\frac{d\sigma_{p\bar{p}}}{d\theta} = \frac{\pi\alpha^{2}\beta C}{2s} \left [|G_M(s)|^{2}(1+cos^{2}\theta) + \frac{4M_p^2}{s}|G_E(s)|^{2}sin^{2}\theta \right],
\label{gegm}
\end{equation}

and the $|G_E/G_M|$ ratio can be extracted from the experimental polar angle distribution.
The ratio depends on energy, also $\frac{4M_p^2}{s}|G_E(s)|^{2}$ is not equal 1, but in spite of this fact  we combine the angular distributions 
from all c.m. energy points in the 1920-2000 MeV interval, because of insufficient statistics. We use  
the procedure described in Sec.~\ref{sec:coll} to obtain the data-MC 
correction for the angular dependence of the detection efficiency. 

\begin{figure}[ptb]
\begin{minipage}[t]{0.49\textwidth}
\centerline{\includegraphics[width=0.99\textwidth,keepaspectratio]{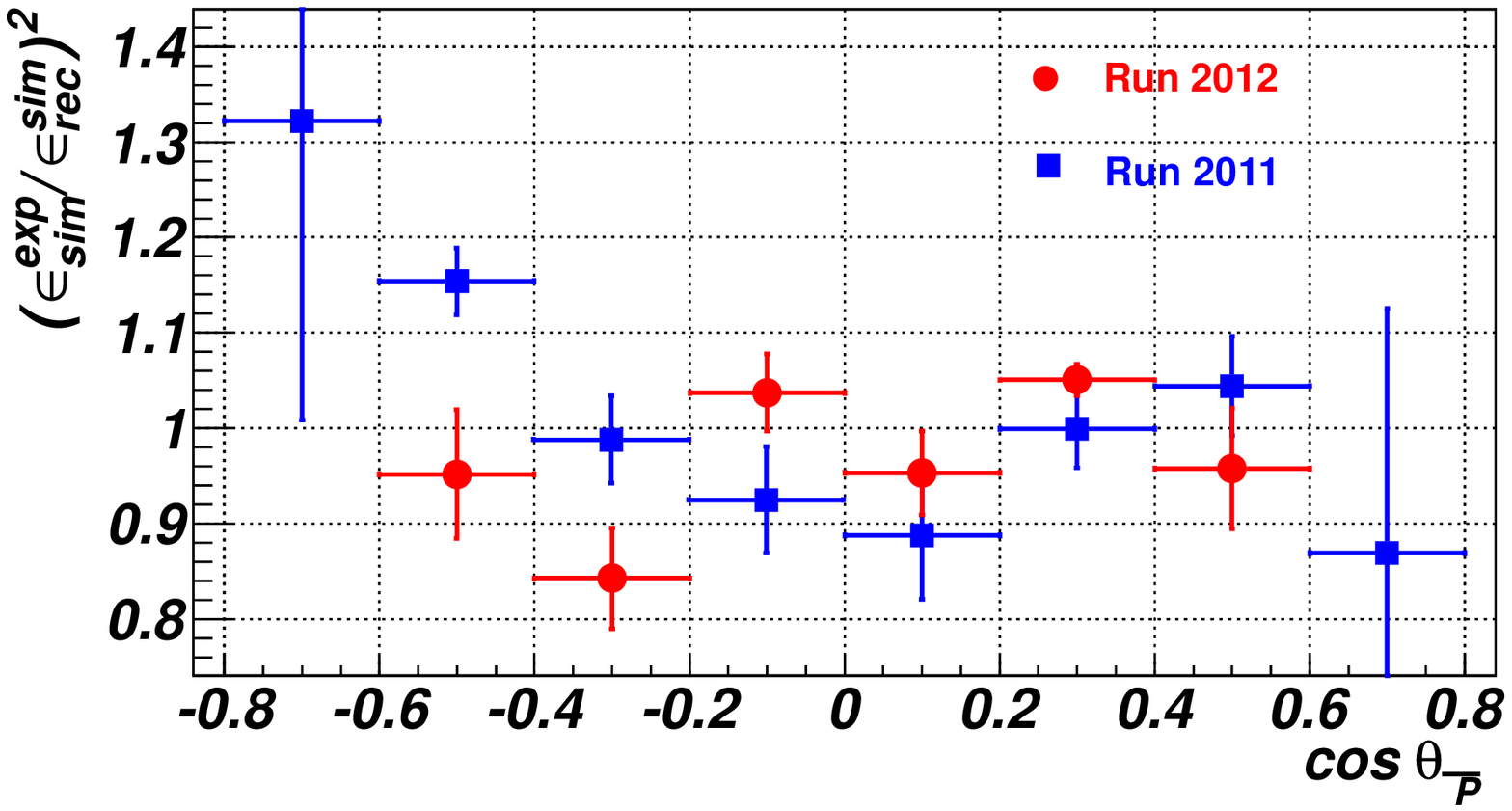}}
\caption{The efficiency correction $(\epsilon_{\rm reg}^{\rm exp}/\epsilon_{\rm reg}^{\rm sim})^2$  vs antiproton polar angle for the  two experimental runs. \label{fig:eff_angle}}
\end{minipage}
\begin{minipage}[t]{0.49\textwidth}
\centerline{\includegraphics[width=0.99\textwidth,keepaspectratio]{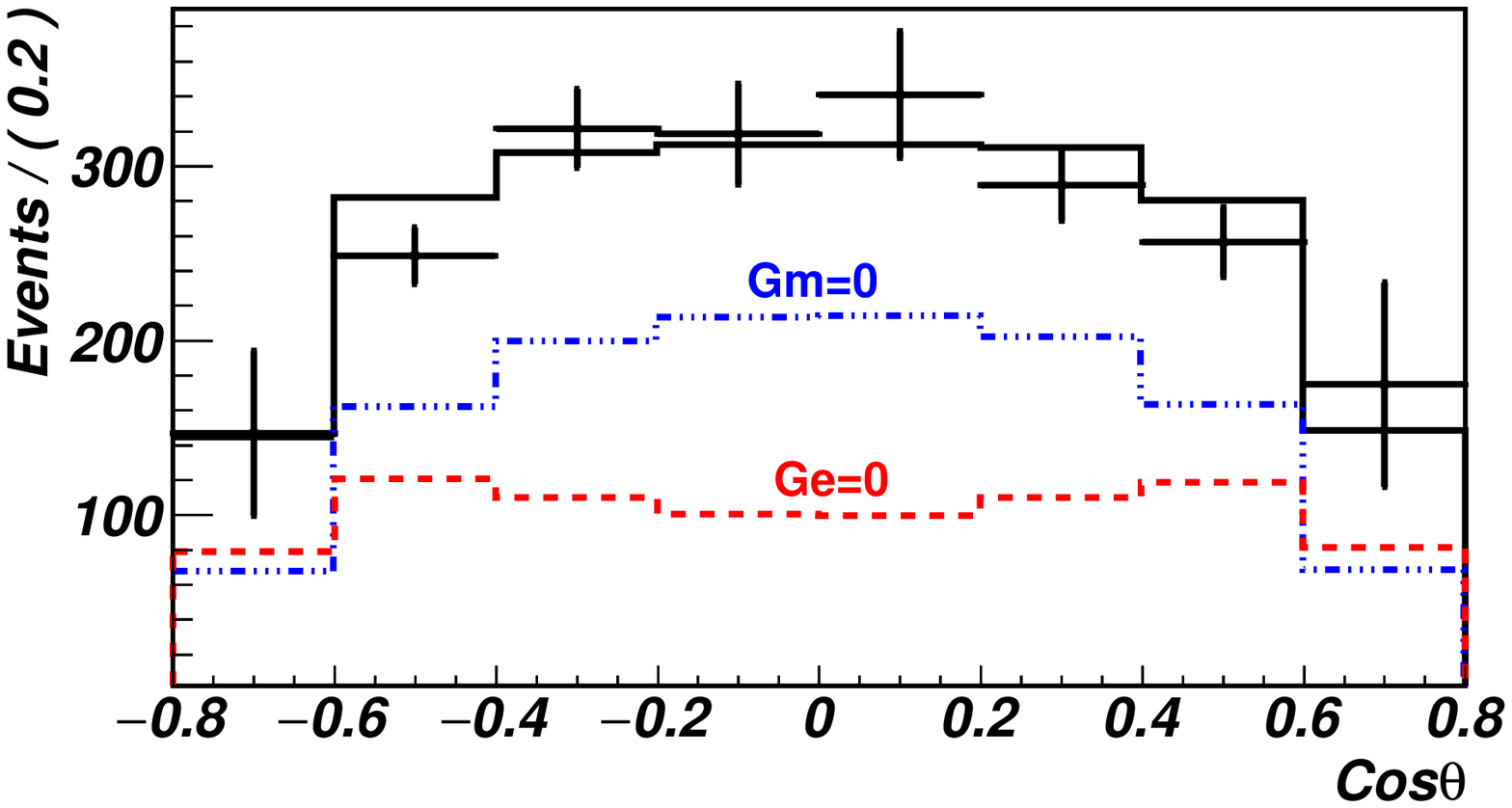}}
\caption{MC-simulated proton polar angle distributions with $G_E$=0 (dashed), $G_M$=0 (dashed-dotted), and fit of experimental data (solid line). \label{fig:gegm_fit}}
\end{minipage}
\end{figure}

These corrections  are shown in Fig.~\ref{fig:eff_angle} for the two 
experimental runs.
We fit the corrected experimental $p\bar{p}$ polar angle distribution shown in Fig.~\ref{fig:gegm_fit} by points with a sum of two functions

\begin{equation}
F^{\rm sim}_{G_{E}=0}+\frac{|G_E|^{2}}{|G_M|^{2}}F^{\rm sim}_{G_{M}=0},
\label{sum_gegm}
\end{equation}

where $F^{\rm sim}_{G_{E}=0}$ and $F^{\rm sim}_{G_{M}=0}$ are contributions to the angular distribution obtained from simulation with $G_{E}=0$ and with $G_{M}=0$, respectively. The numbers of simulated events for  $G_{M}=0$ and $G_{E}=0$ are normalized to the integral of the $(1+cos^2\theta)$ and $sin^{2}\theta$ functions, respectively. 
The fit yields $|G_E/G_M| = 1.49\pm0.23$.

We estimate a systematic error on this value as 20\%, mostly coming from the large statistical errors of the angular correction to the efficiency of Fig.\ref{fig:eff_angle}, and also strong dependence of the corrections on $E_{\rm c.m.}$. 

\begin{figure}[ptb]

\begin{minipage}[t]{0.45\textwidth}
\centerline{\includegraphics[width=0.99\textwidth,keepaspectratio]{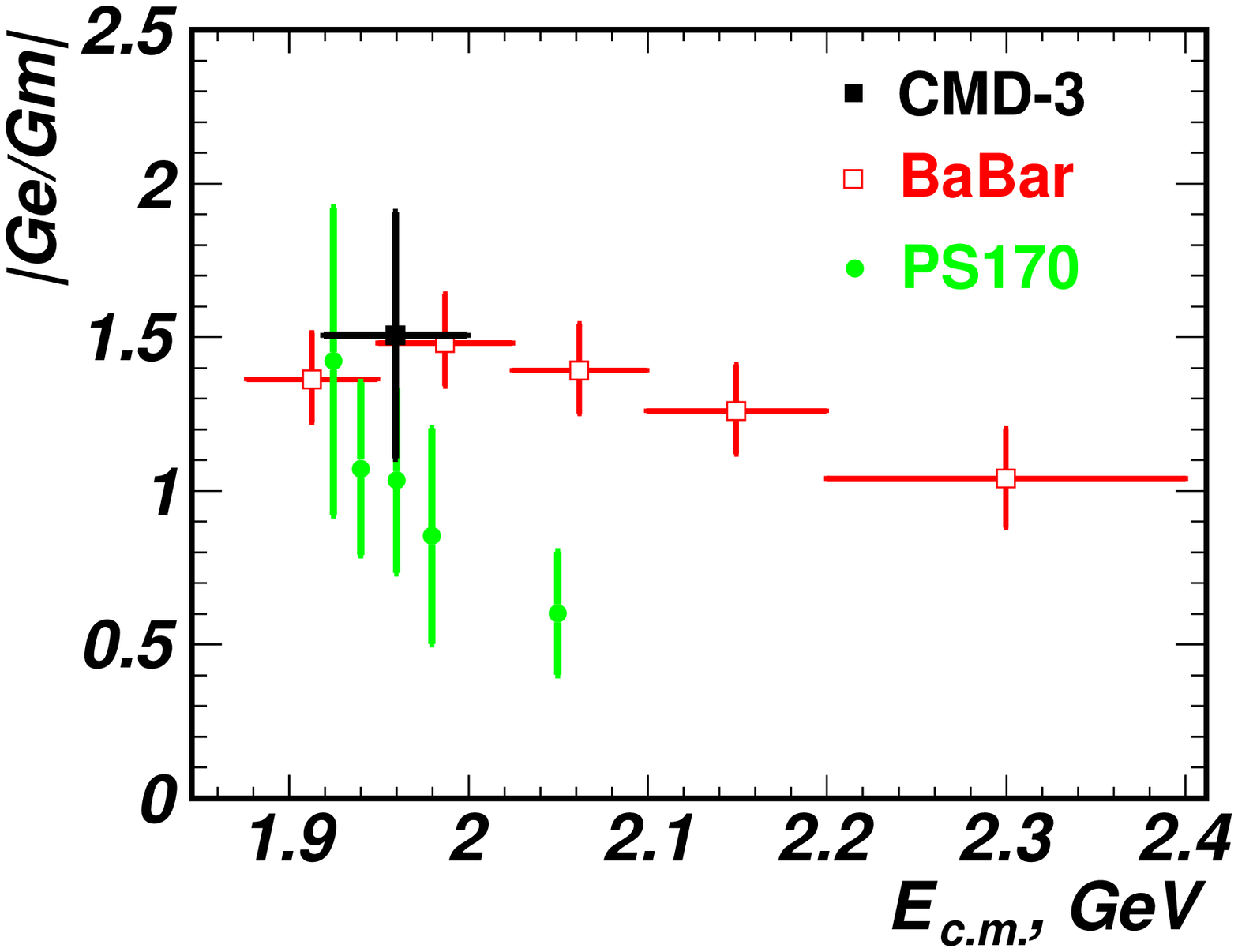}}
\caption{The $|$Ge/Gm$|$ ratio found in this work in comparison with other experiments. \label{fig:gegm}}
\end{minipage}\hspace{0.5cm}
\begin{minipage}[t]{0.45\textwidth}
\centerline{\includegraphics[width=0.99\textwidth,keepaspectratio]{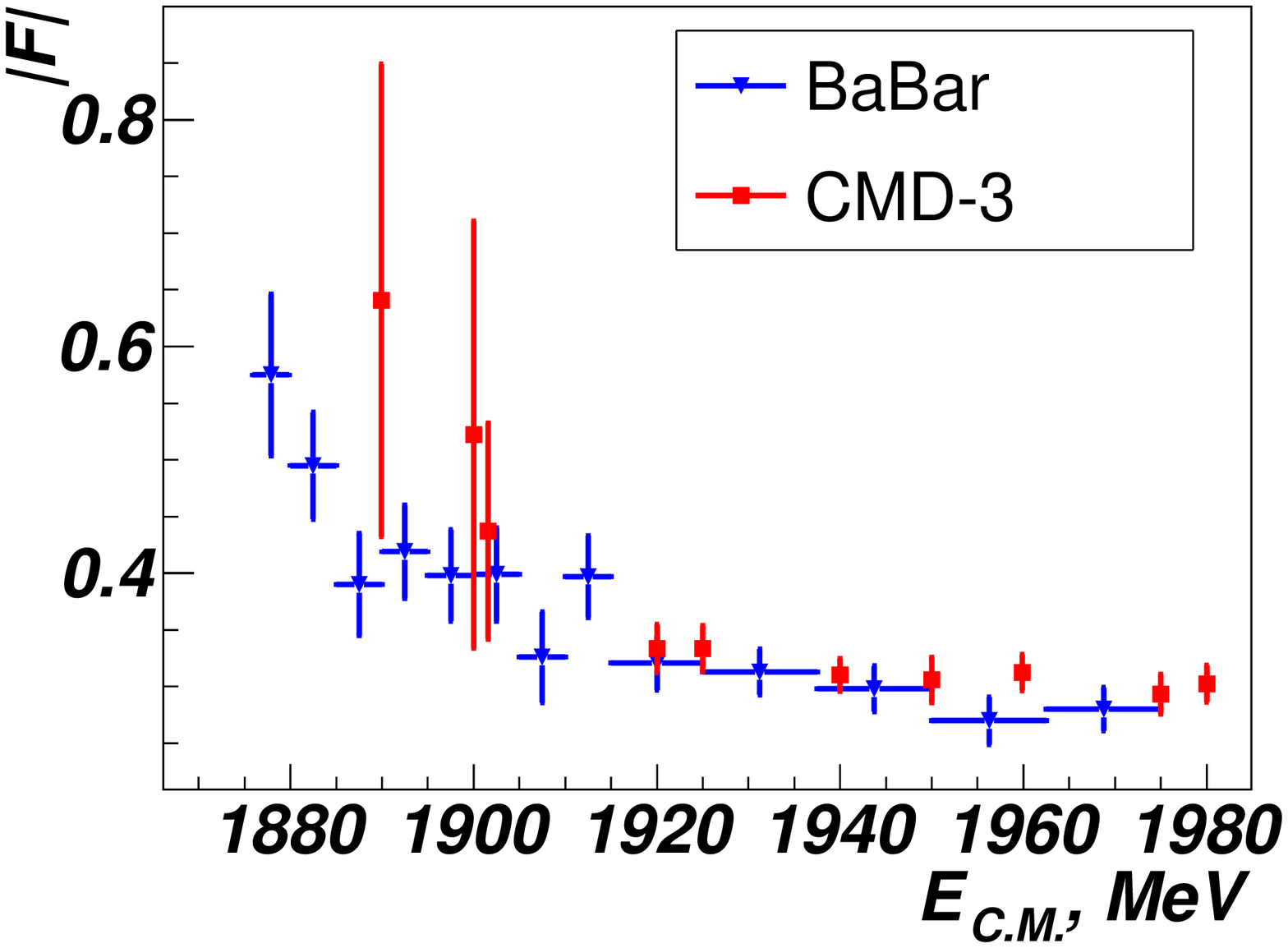}}
\caption{The proton effective form factor measured in this work and in the BaBar experiment~\cite{babar}. \label{fig:ff}}
\end{minipage}

\end{figure}

A comparison of the  measured $|G_E/G_M|$ value with other experiments is shown in Fig.~\ref{fig:gegm} 

\section{Systematic errors}
\subsection{Systematic errors; $p\bar p$ are detected}

The main sources of the systematic errors in this energy range are:

- accuracy of the $\left(\frac{\epsilon_{\rm reg}^{\rm sim}}{\epsilon_{\rm reg}^{\rm exp}}\right)^{2}$ ratio accuracy - 5\%;

- variation of the $|G_E/G_M|$ ratio within error bars leads to 3\% changes in the detection efficiency;

- selection criteria - 2\%;

- luminosity determination - 1\%~\cite{lumin};

- radiative corrections - 1\%~\cite{Sibid};

- beam energy determination accuracy - less than 0.5\% above 955 MeV and 2\% at energy point 950 MeV (run 2012);

- uncertainty of the beam pipe thickness - 2\% above 955 MeV and 10\% at 
950 MeV (run 2012);

Combining above numbers, we estimate a total systematic error as 6\% for beam energies above 955 MeV and 12\% for the energy point 950 MeV(run 2012).

\subsubsection{Systematic errors; $\bar p$ annihilates }

Main sources of systematic errors in this energy range are:

- accuracy of antiproton detection efficiency - 32\%, including uncertainty on the beam energy determination and uncertainty in the beam pipe thickness ;

- $|G_E/G_M|$ accuracy gives 8\% errors in the number of antiprotons stopped in the beam pipe and the DC inner shell. This contribution is considered in antiproton detection efficiency;

- luminosity determination - 1 \%;

- radiative correction - 1 \%;

- selection criteria - 2 \%;

The total systematic errors are 33\%.

\begin{figure}[ptb]
\begin{minipage}[t]{0.99\textwidth}
\centerline{\includegraphics[width=0.99\textwidth,keepaspectratio]{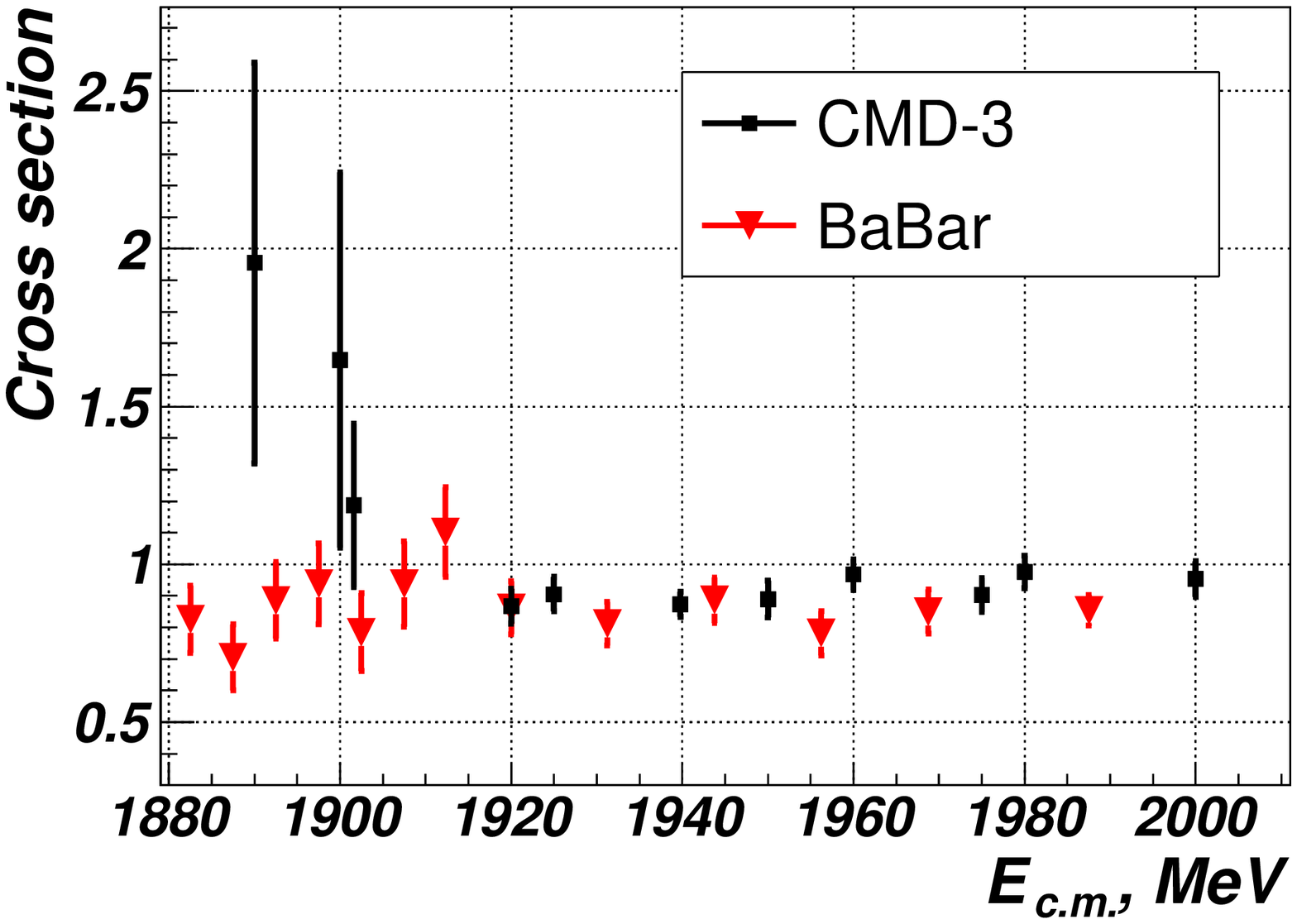}}
\caption{The $e^+ e^- \to p\bar{p}$ cross section measured in this work (circles) in comparison with the  BaBar~\cite{babar} data (triangles). Only statistical errors are shown. \label{fig:cross}}
\end{minipage}
\end{figure}

The  measured cross sections
are shown in Fig.~\ref{fig:ff} and Fig.~\ref{fig:cross}, respectively. The cross sections are listed in Table~\ref{table:sec_above} and Table~\ref{table:sec_below}.

\begin{table}[pt]
  {
    \caption{The c.m. energy, beam energy shift, luminosity, number of selected $e^+e^- \to p\bar{p}$ events, detection efficiency, radiative correction, 
and cross section with statistical and systematic errors. The data for collinear type events.}

    \begin{tabular}{|c|c|c|c|c|c|c|}
      \hline
      $E_{\rm c.m.}$,MeV& $E_{\rm beam}^{\rm shift}$,MeV & L, nb$^{-1}$  &$N_{p\bar{p}}$  &$\epsilon$  &(1-$\delta$)  &$\sigma$, nb           \\ \hline
      1900 (2012) &  0.8$\pm$0.1                  & 900.0        & 164$\pm$13   & $0.2\pm0.04$ & 0.75       &1.2$\pm$0.26$\pm$0.14 \\  \hline
      1920        &  3.3$\pm$0.2                  & 566.9        & 251$\pm$16   & 0.631        & 0.81         &0.87$\pm$0.05$\pm$0.05 \\ \hline
      1925        &  0.5$\pm$0.3                  & 590.8        & 280$\pm$17   & 0.638        & 0.82         &0.90$\pm$0.05$\pm$0.05 \\ \hline
      1940        &  2.4$\pm$0.4                  & 993.8        & 488$\pm$22   & 0.669        & 0.85         &0.87$\pm$0.05$\pm$0.05 \\  \hline
      1950        &  1.2$\pm$0.3                  & 451.0        & 238$\pm$16   & 0.692        & 0.86         &0.89$\pm$0.06$\pm$0.05 \\  \hline
      1960        &  3.0$\pm$0.3                  & 692.2        & 397$\pm$20   & 0.685        & 0.87         &0.96$\pm$0.06$\pm$0.06 \\  \hline
      1975        &  1.3$\pm$0.3                  & 506.6        & 283$\pm$17   & 0.708        & 0.88         &0.90$\pm$0.05$\pm$0.05 \\ \hline
      1980        &  3.6$\pm$0.5                  & 600.6        & 356$\pm$19   & 0.693        & 0.88         &0.98$\pm$0.05$\pm$0.06 \\  \hline
      2000        &  2.3$\pm$0.4                  & 478.0        & 284$\pm$17   & 0.708        & 0.88         &0.95$\pm$0.06$\pm$0.06 \\  \hline
      Total       &                               & 5770.        & 2741$\pm$52  & ---          & ---          & ---                     \\  \hline
    \end{tabular}
    \label{table:sec_above}
  }
\end{table}

\begin{table}[pt]
  {
    \caption{The c.m. energy, luminosity, number of signal events, fraction of antiprotons stopped in beam pipe and DC inner shell, efficiency, cross section 
with statistical and systematic errors, for annihilation events.} 

\small
    \begin{tabular}{|c|c|c|c|c|c|c|}
      \hline
      $E_{\rm c.m.}$,MeV&$E_{\rm beam}^{\rm shift}$ &L, nb$^{-1}$  &$N_{p\bar{p}}$  &(1-$\delta$) &$\epsilon_{\rm stopped}\epsilon_{\rm ann}$        &$\sigma$, nb       \\ \hline
      1890            &     $0.0\pm0.5$      &  527.1      & 79.4$\pm$11  & 0.69        & 0.110            & 1.98 $\pm$0.27 $\pm$0.66\\ \hline
      1900 (2011)     &     $0.0\pm0.5$      &  498.5      & 41.3$\pm$8.5 & 0.75        & 0.067            & 1.65 $\pm$0.34 $\pm$0.54\\ \hline
      1900 (2012)     &     $0.8\pm0.1$      &  900.0      & 44.8$\pm$9.2 & 0.75        & 0.055            & 1.21 $\pm$0.25 $\pm$0.40\\ \hline
    \end{tabular}
    \label{table:sec_below}
  }

\end{table}

\section{ \boldmath CONCLUSION}
\hspace*{\parindent}
The $e^+e^- \to  p\bar{p}$ cross section has been measured using a data sample of 6.8 pb$^{-1}$ collected in the center-of-mass energy range from $p\bar{p}$ threshold to 2 GeV.  Results agree with the previous BaBar experiment 
and have comparable or better statistical and systematic errors. 
The value of the ratio $|G_E/G_M| = 1.49 \pm 0.23\pm 0.30$ has been found 
in the energy range from 1.92 to 2 GeV, and is in agreement with BaBar data. The expected tenfold increase in the luminosity of VEPP-2000 will allow to measure the proton form factor and $|G_E/G_M|$ ratio with much better accuracy.

\subsection*{ \boldmath ACKNOWLEDGEMENTS}
\hspace*{\parindent}
This work is supported  in part by the Russian Education and Science Ministry, by the Russian Foundation for
  Basic Research grants

RFBR 14-02-31478,
RFBR 14-02-00580-a,
RFBR 14-02-00047-a,
RFBR 14-02-31275-mol-a
RFBR 13-02-00215-a,
RFBR 13-02-01134-a.

\end{document}